\documentclass{appolb}

\usepackage{amsmath}
\usepackage{amsfonts}
\usepackage{amssymb}
\usepackage{graphicx}

\def\half{\textstyle{\frac{1}{2}}}

\begin{document}

\title{Fractal properties
of the diffusion coefficient\\ in a simple deterministic dynamical
system:\\ a numerical study }

\author{Zbigniew Koza\thanks{Electronic address: \texttt{zkoza@ift.uni.wroc.pl}}
 \address{
             Institute of Theoretical Physics, University of
             Wroc{\l}aw,\\ pl.\ Maxa Borna 9, 50204 Wroc{\l}aw,
             Poland
           }
 }


 \maketitle

\begin{abstract}
Using a numerical library for arbitrary precision arithmetic I study the
irregular dependence of the diffusion coefficient on the slope of a
piecewise linear map defining a dynamical system. I find that the graph of
the diffusion coefficient as a function of the slope has the fractal
dimension 1, but the convergence to this limit is slowed down by logarithmic
corrections. The exponent controlling this correction depends on the slope
and is either 1 or 2 depending on existence and properties of a Markov
partition.
\end{abstract}



\section{Introduction}
\label{Introduction}
Determining the transport coefficients of many-particle systems is one
of the fundamental problems of nonequilibrium statistical physics.
This is also a notoriously difficult problem: it turns out that even
in a simplified system where a single particle moves in a periodic
array of scatterers, the drift and diffusion coefficients are highly
irregular, apparently nowhere differentiable functions of control
parameters \cite{KlagesPHD,KlagesPRE99,DorfmanBook,KlagesHab} (an
example is also shown in Fig.~\ref{Fig1} below). Closer inspection of
these functions reveals that usually their ``irregularities'' are not
random, but rather form patterns. This has led some researchers to the
idea that the graphs of these functions are fractals with nontrivial,
perhaps locally varying fractal dimension. This is an interesting
concept, since fractal structures are quite commonly found in
dynamical systems in various contexts \cite{DorfmanBook,KlagesHab}.

Until very recently no reliable investigation of fractal properties of
transport coefficients was possible because all general methods of
calculating transport coefficients in dynamical systems -- e.g.\ the
transition matrix technique combined with the escape rate formalism
\cite{KlagesPHD,KlagesPRE99,DorfmanBook,KlagesHab,Gaspard90,Gaspard92},
the Green-Kubo formula \cite{KlagesPHD,DorfmanBook,KlagesHab}, or the
periodic-orbit formalism \cite{DorfmanBook,CvitanovicWWW} --
eventually lead to complicated and time-consuming numerical
calculations. Moreover, usually they are applicable only for some
special values of the control parameters. For these reasons none of
them could be used to collect a sufficiently large number (counted at
least in millions) of very precise data required in fractal analysis.
This situation changed when Groeneveld and Klages \cite{Groeneveld}
gave exact formulas for the transport coefficients in a simple
one-dimensional dynamical system introduced by Grossmann and Fujisaka
\cite{Grossman82}.

The system investigated by Groeneveld and Klages can be considered as a
model of a particle moving in a one-dimensional array of scatterers. The
role of the equation of motion is played by a one-dimensional map $M$
\begin{equation}
  x_{n+1} = M_{a,b}(x_n)
\end{equation}
where $n$ is a discrete-time variable and the map $M_{a,b}\colon \mathbf{R}
\to \mathbf{R}$ is given by a simple linear function
\begin{equation}
  M_{a,b}(x) = ax+b,\quad x \in I_0^-
\end{equation}
on the fundamental interval $I_0^- = [-\half,\half)$, with $a>1, b \in
\mathbf{R}$ being the control parameters representing the slope and the bias
of the map, respectively; the map is then continued periodically onto the
real line by a lift of degree one, i.e. by requiring that
\begin{equation}
  M_{a,b}(x+1) = M_{a,b}(x) + 1,\quad x \in \mathbf{R}.
\end{equation}
For $a>1$ the Lyapunov exponent of this system is positive, and so the
dynamics defined by $M_{a,b}$ is chaotic. If we choose an arbitrary
number $x_0$ as the starting point, the resulting sequence $(x_n)$
will almost always look ``random'' (the set of points $x_0$ generating
a regular, periodic or quasi-periodic sequence $(x_n)$ is of Lebesgue
measure 0). For suitably chosen values of $a$ and $b$ the
deterministic dynamics defined by the map $M_{a,b}$ is equivalent to
a  Markov stochastic process of random-walk type: each deterministic
trajectory $(x_n)$ is equivalent to some particular realization of the
corresponding random-walk process, and taking the average over all
initial states is equivalent to calculating averages over the
corresponding Gibbs ensamble. Actually any random-walk Markov process
in a one-dimensional periodic system with fixed transition rates can
be translated into the language of simple piecewise linear
deterministic maps \cite{Claes93}. From this point of view there is no
surprise that the process defined by $M_{a,b}$ (or similar maps) is
called ``deterministic diffusion'' and that the two basic transport
coefficients, the drift velocity $J$ and diffusion constant $D$, can be
defined as
\begin{equation}
  J = \lim_{n\to\infty} \frac{\langle x_n\rangle}{n}, \qquad
  D = \lim_{n\to\infty} \frac{\langle x_n^2\rangle - \langle x_n \rangle^2}{2n}
\end{equation}
where $\langle\cdots\rangle$ denotes the average over the uniform ensamble of
initial values $x_0$.

The graph of the diffusion coefficient $D$ for the map $M_{a,b}$ as a
function of the slope $a$ for the bias $b=0$ is shown in Fig.~\ref{Fig1}. The
fractal properties of this highly irregular graph (as well as that of the
drift velocity $J$) were recently studied by Klages and Klau\ss\
\cite{Klages03}. Using two numerical methods: the box counting and the
autocorrelation function methods, they found that the local fractal
dimensions of graphs of $J$ and $D$ are well-defined, but highly irregular
functions of $a$ and $b$. In other words they found that the graph shown in
Fig.~\ref{Fig1} cannot  be described with a single fractal dimension, but
rather by a set of quickly varying local fractal dimensions. Taking this
into account they suggested that the local fractal dimensions of graphs of
$J$ and $D$ as functions of the slope $a$ with the bias $b$ fixed are
fractal themselves. This, in turn, leads to the concept of a ``fractal
fractal dimension of deterministic transport coefficients'' \cite{Klages03}.
\begin{figure}
\begin{center}
  \includegraphics[width=0.5\textwidth, clip=true]{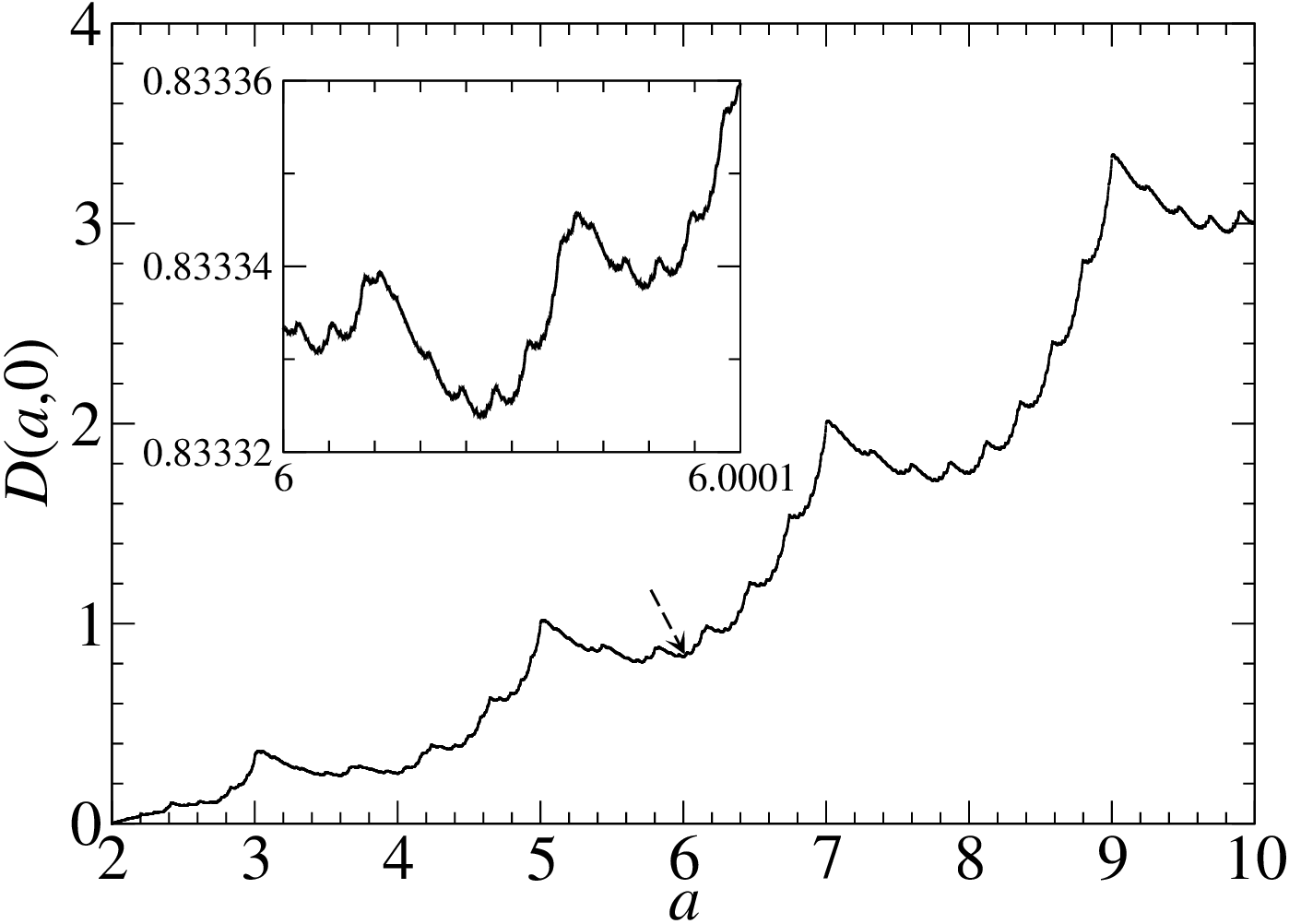}
  \caption{
    \label{Fig1}
   The diffusion coefficient $D$ for the map $M_{a,b}$
   as a function of the slope $a$ for the bias
   $b=0$. The inset depicts a blow-up of a region $6 \le a \le 6.0001$
   (pointed at by the arrow).}
\end{center}
\end{figure}

Klages and Klau\ss\ found the local fractal dimension $\mathcal{D}$ of
$J(a,b)$ and $D(a,b)$ to be very close to 1. The autocorrelation function
method gave $1.05  \lesssim\mathcal{D} \lesssim 1.15$, and the box-counting
method gave even lower values $1.02 \lesssim \mathcal {D} \lesssim 1.05$.
All these values are very close to 1, i.e.\  the fractal dimension of a
regular, ``smooth'' curve. This suggests that perhaps  $\mathcal{D} = 1$ and
that the results obtained in the above-mentioned computer simulations
reflected extremely slow convergence of $\mathcal{D}$ to its true asymptotic
limit of infinitesimally small ``boxes''. Such slow convergence is often
caused by logarithmic corrections. The main purpose of my paper is thus to
try and find such tiny corrections. To accomplish this task I will use a
different method of calculating the local fractal dimension of a curve -- the
so called oscillation method \cite{Tricot} -- and I will carry out all
calculations with a help of a special high-accuracy numerical library, which
will enable me to consider exceptionally small ``boxes''. For sake of
simplicity I shall restrict the present study to the simplest case of zero
bias ($b=0$) where, due to symmetry, the drift velocity $J = 0$.

The structure of the paper is as follows. Section \ref{SectionMethod}
briefly describes the method I have used to determine the local fractal
dimension. It describes all steps necessary to calculate the diffusion
coefficient for the map $M_{a,b}$, a method of calculating the local fractal
dimension of a continuous curve (``the oscillation method''), mathematical
formulation of the main conjecture about the logarithmic convergence of the
fractal dimension to its limiting value, and the numerical aspects of the
algorithms used. Section \ref{SectionResults} presents the main results.
Finally, section \ref{SectionConclusions} is devoted to discussion of
results.

\section{Method} \label{SectionMethod}
\subsection{Transport coefficient}
The explicit formulas \cite{Groeneveld} for the transport coefficients in the
model depend on some auxiliary variables. For each $a>1, b\in \mathbf{R}$,
and $\epsilon = \pm$ we define two infinite sequences, $(y_r^\epsilon)$, $r
= 0,1,\ldots$, and $(n_r^\epsilon)$, $r = 1,2,\ldots$, consisting of real
and integer numbers, respectively. Their values are uniquely determined by
demanding that $y_0^\epsilon = \frac{\epsilon}{2}$ and that for each $r> 0$
\begin{equation}
\label{2Reccursion}
n_r^\epsilon + y_r^\epsilon = a y^\epsilon_{r-1} + b,
\end{equation}
with additional conditions $n_r^\epsilon \in \mathbb{Z}$ and
$y_r^\epsilon\in  I^\epsilon_0$, where $I_0^+ \equiv \left( -\half, \half
\right]$  and $I_0^- \equiv \left[ -\half, \half
\right)$. Next we define ``N-numbers'':
\begin{equation}
\label{Nr}
  N_r^\epsilon = -\frac{\epsilon}{2} + \sum_{s=1}^r n_s^\epsilon,
\end{equation}
\begin{equation}
\label{Nkl:eps}
  N_{k,l} = \frac{1}{k!l!}\sum_{r=0}^\infty a^{-r}[(N_r^+)^k - (N_r^-)^k]
  r^l,
\end{equation}
where $r,k,l \ge 0$. The basic transport coefficients,  $J$ and $D$,  can be
now expressed (\cite{Groeneveld}, cf \cite{Koza99}) as
\begin{equation}
  \label{JD}
  J = \frac{N_{2,0}}{N_{1,1}},\quad
  D = \frac{N_{3,0} - N_{2,1}J + N_{1,2} J^2}{N_{1,1}}.
\end{equation}

\subsection{The oscillation method}
The fractal dimension $\mathcal{D}$ of a continuous function $f$ defined on
an interval $[t_0,t_1]$ can be evaluated through analysis of its H\"older
exponents \cite{Tricot} and $\tau$-oscillations \cite{Tricot}. The
$\tau$-oscillation of $f$ at $t\in (t_0,t_1)$ is defined as
\begin{equation}
  \label{def-osc}
  \text{osc}_\tau(f; t) =
   \sup_{|t-t'|\le \tau}f(t') - \inf_{|t-t'|\le \tau}f(t').
\end{equation}
If there exist constants $c>0$ and $0 < H \le 1$ such that for all $\tau$
\begin{equation}
  \text{osc}_\tau(t) \le c\tau^H
\end{equation}
then $f$ is called a Holderian of exponent $H$ at $t$ and its fractal
dimension $\mathcal{D}$ at $t$ is related to $H$ through
\begin{equation}
 \label{DeltaLE}
  \mathcal{D} \le 2-H.
\end{equation}
Similarly, if there exist constants $c>0$ and $0 < H \le 1$ such that for all
$\tau$
\begin{equation}
  \text{osc}_\tau(t) \ge c\tau^H
\end{equation}
then $f$ is called an anti-Holderian of exponent $H$ at $t$ and
\begin{equation}
 \label{DeltaGE}
  \mathcal{D} \ge 2-H.
\end{equation}

\subsection{Reformulation of the problem and numerical implementation}
We are now ready to formulate our main conjecture: for the map $M_{a,b}$ with
$b=0$
\begin{equation}
 \label{conjecture}
  \frac{\text{osc}_\tau(D;a)}{\tau} \approx
  c(a)\left[-\log(\tau)\right]^{\gamma(a)},\quad \mathrm{as~} \tau \to 0, 
\end{equation}
where the prefactor $c(a)>0$ and the exponent $\gamma(a) \ge 0$. Owing to
(\ref{def-osc}) -- (\ref{DeltaGE}) this implies that the fractal dimension
of the graph of $D$ as a function of the control parameter $a$ is equal 1,
but the convergence to this limiting value is logarithmically slow, with the
exponent $\gamma(t)$ controlling the convergence rate.

I checked conjecture (\ref{conjecture}) using equations (\ref{2Reccursion})
-- (\ref{JD}). Because the consecutive terms of the sequences $n_s^\pm$ in
(\ref{Nr}) can be generated extremely efficiently (it took less than a second
to generate 20~000 data points used to draw Figure~\ref{Fig1}), I decided to
employ GMP (GNU multiple precision arithmetic library, version 4.1.2) -- a
library for arbitrary precision arithmetic \cite{GMP}. Although any
calculations performed by such a library must be several orders of magnitude
slower that those performed directly, thanks to the GMP I was able to study
numerically the limit of $\tau \to 0$ for $\tau$ ranging from $1$ down to at
least $10^{-100}$ (the accuracy of typical computers currently available,
without using such special-purpose libraries, is limited typically to about
$10^{-17}$). Such fine resolution of my calculations will turn out crucial
for determining logarithmic corrections predicted by formula
(\ref{conjecture}).

\section{Results}
\label{SectionResults}

In my present study  I decided to concentrate on verifying conjecture
(\ref{conjecture}) for the symmetrical case of vanishing bias ($b =
0$) and for several selected values of the slope $a$. I decided to
examine especially carefully the slopes corresponding to finite Markov
partitions. These values are particularly interesting from the
theoretical point of view \cite{KlagesPHD,KlagesPRE99,DorfmanBook};
for example, even though the diffusion coefficient $D$ is always
continuous in $a$ \cite{Groeneveld}, the numerator and denominator in
(\ref{Nkl:eps}) can be discontinuous for such (and only such) slopes.
One of the reasons why I chose to set $b=0$ is that in this case the
slopes corresponding to Markov partitions can be found easily on a
computer: they are algebraic numbers generating periodic sequences
$n_k$ and $y_k$. Such values of $a$ will be henceforth referred to as
``Markov slopes''.

\begin{figure}[b]
  \includegraphics[width=\textwidth, clip=true]{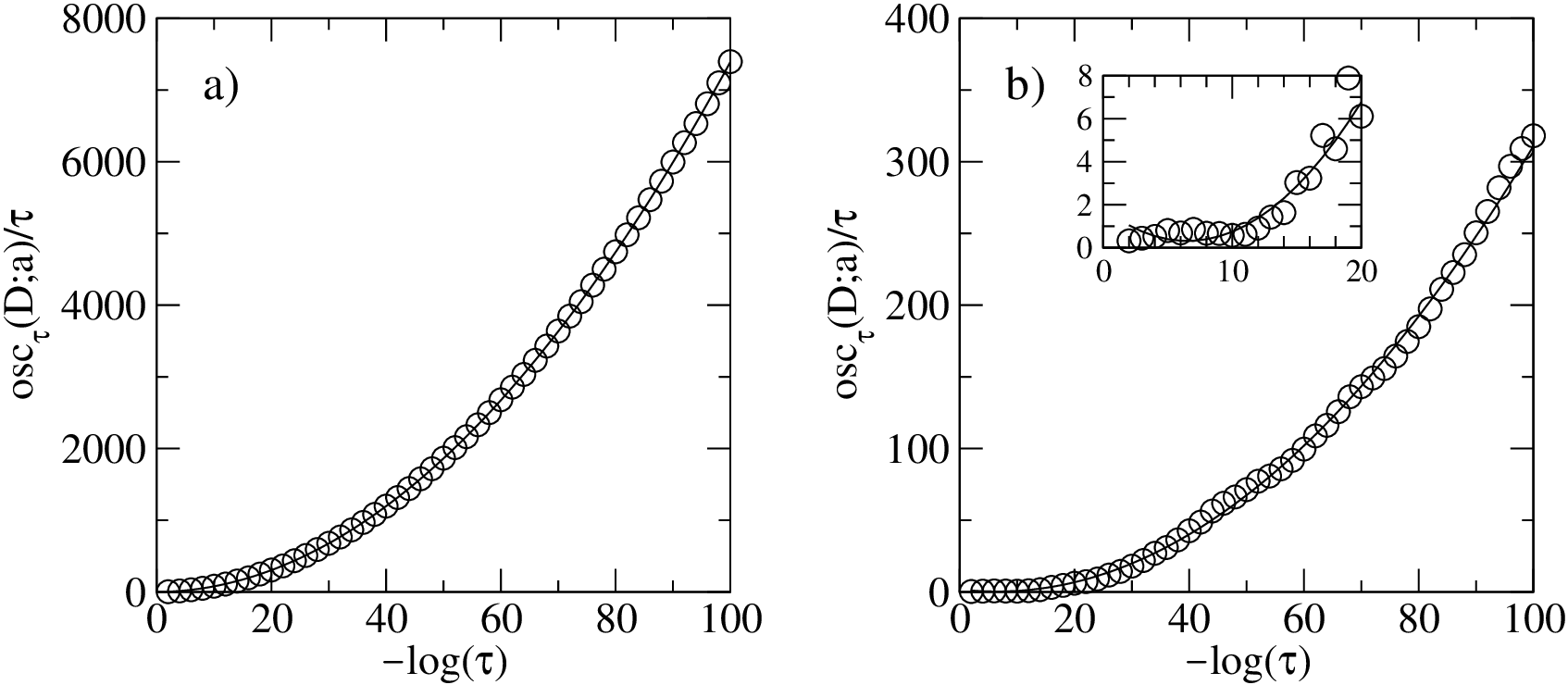}
  \caption{
    \label{Fig2}
   Oscillations of the diffusion coefficient rescaled in accordance with eq.\
   (\ref{conjecture})
   calculated on intervals $(a-\half\tau,a+\half\tau)$ for $\tau =
   10^{-n}, n = 1,\ldots,100$ and for (a)~$a = 3$ and (b) $a \approx
   3.314$.
   The circles represent results calculated from conjecture (\ref{conjecture}),
   and the solid lines are quadratic fits.
   The inset in graph (b) shows the blow-up of results obtained for $1 > \tau
   \ge 10^{-20}$.
 }
\end{figure}

Figure \ref{Fig2} presents results obtained for two Markov slopes generating
strictly periodic sequences (i.e., $y^\epsilon_{k+l} = y^\epsilon_k$ for a
period $l>0$ and all $k\ge0$). The value used in Fig.\ \ref{Fig2}a is $a=3$.
This slope generates simple sequences $y^+_k$ and $y^-_k$ of period 1. The
oscillations of the diffusion coefficient on intervals
$(3-\half\tau,3+\half\tau)$, rescaled by $\tau$, obtained for $\tau =
10^{-n}, n = 1\ldots100$, are represented in this plot by circles, while the
solid line represents a quadratic fit of form $px^2 + qx + r$ with
$p=0.733$, $q=0.57$, and $r=-0.2$. The fit is excellent, which suggests that
for this slope the exponent $\gamma$, as defined in eq.\ (\ref{conjecture}),
is~2. I have obtained similarly good quadratic fits for other integer odd
slopes, which also generate simple sequences of period 1 \cite{KlagesPRE99}.

A question arises whether the length of the period has any influence on the
limiting value of $\mbox{osc}_\tau(\mathcal{D};a)/\tau$. Figure~\ref{Fig2}b
shows the results obtained for the largest root of the polynomial $a^4 - 4a^3
+ 2a^2 + 3$, i.e.\ for $a\approx 3.314$. For this slope the period $l=4$.
Just as in the previous example, $\mbox{osc}_\tau(\mathcal{D};a)/\tau$ can be
approximated by a quadratic, although the fit is not as excellent as for
$a=3$. The inset in this figure depicts the blow-up of the data obtained for
$1 > \tau \ge 10^{-20}$. It shows that for $1 > \tau > 10^{-10}$ the value of
$\mbox{osc}_\tau(\mathcal{D};a)/\tau$ fluctuates about a constant value,
which might suggest that the exponent $\gamma$ vanishes. This example
demonstrates that the ``resolution'' of calculations used in ref.\
\cite{Klages03}, e.i.\ $\tau_{\mathrm{min}} \lesssim 10^{-10}$, is
insufficient to find the true asymptotic fractal properties of the graph of
the diffusion coefficient as a function of the slope $a$.

\begin{figure}[b]
  \includegraphics[width=\textwidth, clip=true]{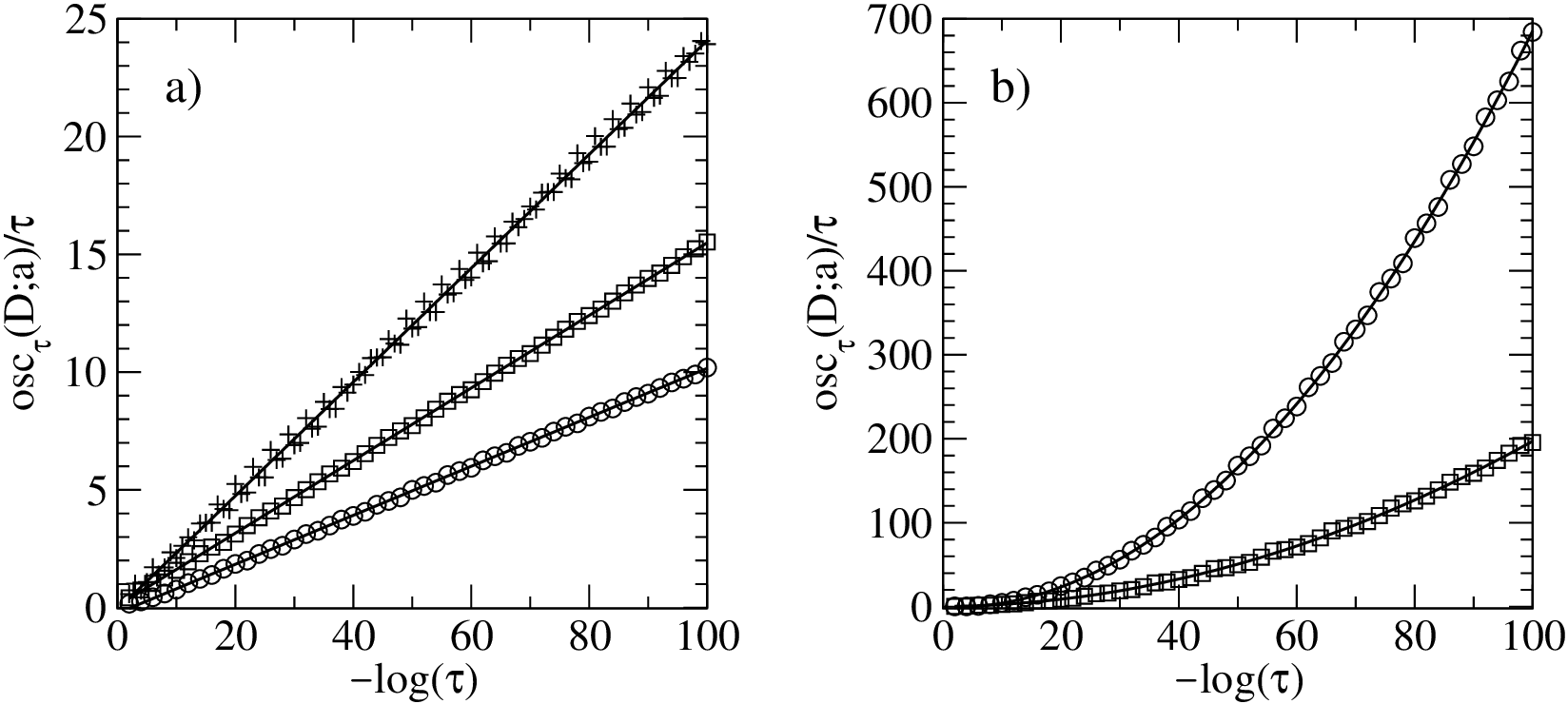}
  \caption{
    \label{Fig3}
   The same as in Fig.\ \protect\ref{Fig2}.
   (a) $a = 4$ (circles), $a \approx 3.3497$ (squares), $a \approx 3.043$ (pluses)
   with linear fits (solid lines);
   (b) $a \approx 3.160$ (circles), $a \approx 4.200$ (squares) with quadratic fits (solid lines).
    See text for details.
 }
\end{figure}

All other Markov slopes generate sequences $n_k^\epsilon$ and $y_k^\epsilon$
with the period starting at a term $m>0$. There are two cases: either all
terms of $y_k^+$ are different from all the terms of $y_k^-$ or the two
sequences have the same period sequence. Let's start from the latter case.
Three examples of results obtained for Markov slopes of this type are shown
in Figure \ref{Fig3}a. The data depicted in this graph were calculated for
$a=4$ (circles), for the largest root of  $a^4 - 4a^3 + 2a^2 +2$, i.e.\ for
$a\approx 3.3497$ (squares), and for the largest root of $a^5 - 4a^4 + 2a^3 +
3a^2 - 2a + 4$, i.e.\ for $a \approx 3.043$ (pluses). For the first two of
the above slopes all the terms of the sequences $y_k^\epsilon$ eventually
vanish. In the case of $a=4$ this happens for $k\ge1$, and for $a\approx
3.3497$ the terms of $y^\epsilon_k$ vanish for $k\ge4$. As for the third
example, $a\approx3.043$, the period starts at $k=2$ and has the length 6.
Specifically, the first five terms of $y_k^+$ are (to 3 S.F.) $1/2$,
$-0.478$, $-0.456$, $-0.387$, $-0.179$, and for $k\ge5$ the terms satisfy
$y_k^+ = -y_{k-3}^+ = y_{k-3}^- $. As we can see, in all these cases
$\mbox{osc}_\tau(\mathcal{D};a)/\tau$  diverges linearly with
$\log(\tau^{-1})$, suggesting that conjecture (\ref{conjecture}) is satisfied
with the exponent $\gamma = 1$.

The results obtained for the last category of Markov slopes, i.e. slopes
generating two disjoint periodic sequences $y_k^+$ and $y_k^-$, are shown in
Fig.~\ref{Fig3}b. The two data sets were collected for the largest root of
$a^5 - 4a^4 + 2a^3 + a^2 + 4a - 2$, i.e.\ for $a\approx 3.16$ (circles) and
for the largest root of $a^5 - 4a^4 - 4a^2 + a + 4$, i.e.\ for $a\approx
4.20$ (squares). The first of these slopes, $a\approx3.16$, generates a
sequence with a period of length 3 starting at $y_2^+$, while $a\approx
4.20$ generates a sequence of period length 4 starting at $y_1^+$. As we can
see, for these Markov slopes $\mbox{osc}_\tau(\mathcal{D};a)/\tau$ can be
very well approximated by a quadratic function of $\log(\tau^{-1})$. This
suggests that in this case conjecture (\ref{conjecture}) is satisfied with
the exponent $\gamma = 2$.

\begin{figure}
  \includegraphics[width=\textwidth, clip=true]{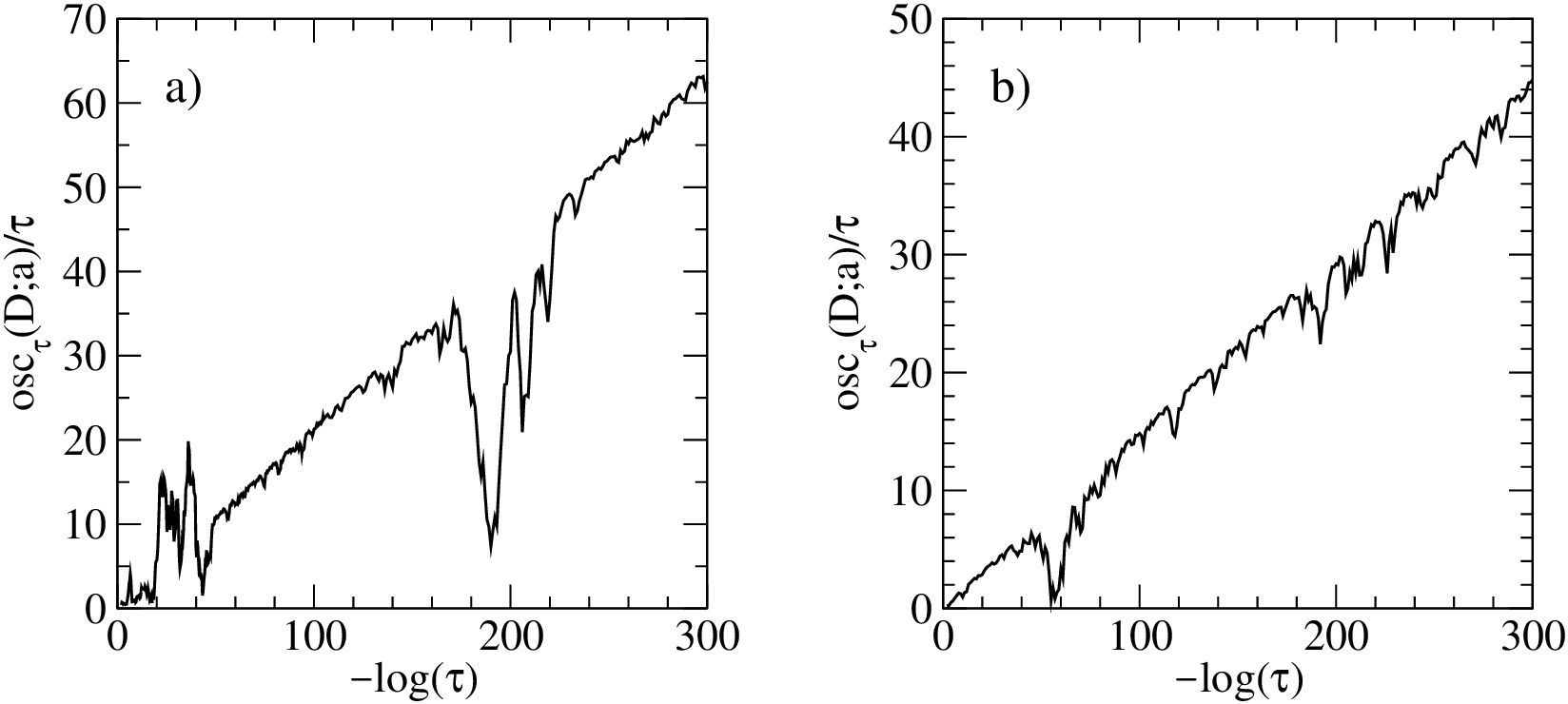}
  \caption{
    \label{Fig4}
   The same as in Fig.\ \protect\ref{Fig2} for two slopes $a$ that do
   not correspond to finite Markov partitions:
   (a) $a = \pi \approx 3.14$ and (b)  $a = 4.5$.
 }
\end{figure}

Figure \ref{Fig4} presents results for two slopes $a$ which do not
correspond to a Markov partition. These are $a = \pi \approx 3.14$
 (left panel) and $a=9/2$
 (right panel). The results turned out to be
``noisy'' and I had to increase the resolution down to $\tau_{\mathrm{min}} =
10^{-300}$. As can be seen, in both cases
$\mbox{osc}_\tau(\mathcal{D};a)/\tau$ has a clear trend linear in
$\log(\tau^{-1})$. However, this limiting behaviour is disturbed by very
large fluctuations. Actually, for $a = \pi$ these fluctuations are so large
that my numerical results cannot rule out the possibility that in this case
the limit $\lim_{\tau\to0} \mbox{osc}_\tau(\mathcal{D};a)/\tau$ does not
exist.

\section{Conclusions}

In my study I proposed and verified numerically a hypothesis that the
local fractal dimension of the graph of the diffusion coefficient of
the piecewise linear map $M_{a,b}$ as a function of the slope $a$ is
$\mathcal{D} = 1$ and that the convergence to this limit is slowed down
by logarithmic corrections described by eq.\ (\ref{conjecture}). This
contradicts the earlier findings of Klages and Klau\ss\
\cite{Klages03} that this graph is a fractal with a locally varying
fractal dimension $\mathcal{D}(a)
> 1$ which, when plotted as a function of the slope, forms a fractal itself.

I found that the exponent $\gamma$, which controls the logarithmic
correction, is actually a function of the slope $a$. Interestingly, $\gamma$
appears to be a discontinuous function that can take only one of two
 values: 1 or 2. The value of 1 corresponds to Markov slopes that
generate two disjoint sequences $y_k^+$ and $y_k^-$, and the value of
$\gamma= 2$ corresponds to Markov slopes that generate periodic sequences
$y_k^+$ and $y_k^-$ with the same period terms. The case of slopes that do
not generate Markov partitions is not clear -- apparently $\gamma = 1$, but
this statement cannot be verified numerically because of very large
fluctuations making the convergence extremely slow. Note that my findings
imply that both of the sets $\{ a\colon \gamma(a) = 1\}$ and $\{ a\colon
\gamma(a) = 2\}$ are dense, and hence that $\gamma(a)$ is nowhere continuous.

Any numerical study is naturally restricted to investigation of several
particular cases. It cannot be ruled out that I have overlooked some
categories of slopes $a$ for which the local fractal dimension of the
graph of the diffusion coefficient behaves in a way not predicted in
this study. Numerical investigation of the logarithmic correction to
the asymptotic limit is here a particularly delicate problem, as
apparently $\gamma$ is a nowhere continuous function of the slope. Only
through an analytical approach could this problem be solved
conclusively. Such a study will be published elsewhere.

\label{SectionConclusions}


\section*{Acknowledgments}

I thank R.\ Klages for many inspiring discussions. Support from the
Polish KBN Grant Nr 2 P03B 030 23 is gratefully acknowledged.



\end{document}